\documentstyle[floats,prd,aps,epsf]{revtex}
\begin{document}
\tighten
\preprint{\vbox{
        \hbox{UTTG-10-95}
    \hbox{MADPH-95-914}
    \hbox{hep-ph/9601298}}
}
\title{Photoproduction of J/$\psi$ in the forward region}
\author{James Amundson\thanks{Email: amundson@pa.msu.edu}}
\address{Department of Physics, University of Wisconsin, Madison, WI 53706}
\address{Current Address: Department of Physics and Astronomy,
 Michigan State University, East Lansing, MI 48824}
\author{Sean Fleming\thanks{Email: fleming@phenxs.physics.wisc.edu}}
\address{Department of Physics, University of Wisconsin, Madison, WI 53706}
\author{Ivan Maksymyk\thanks{Email: maksymyk@alph02.triumf.ca}}
\address{Theory Group, Department of Physics, University of Texas,
         Austin, TX 78712}
\address{Current Address:
TRIUMF, 4004 Wesbrook Mall, Vancouver, BC, Canada, V6T 2A3}
\maketitle
\begin{abstract}
We study the phenomenology of fixed-target elastic $J/\psi$
photoproduction in the NRQCD factorization formalism.  Our the goal
is to test an essential feature of this formalism --- the color-octet
mechanism. We obtain an order-of-magnitude estimate for a certain
linear combination of NRQCD color-octet matrix elements. Our estimate
is consistent with other empirical determinations and with the
$v$-scaling rules of NRQCD.
\end{abstract}


\section{Introduction}

We study the phenomenology of the NRQCD factorization formalism,
which is a framework for writing inclusive rates of production and
decay for quarkonium.  Specifically, we consider photoproduction of
$J/\psi$.  This process probes an essential feature of the NRQCD
formalism --- the octet-mechanism. We approach our subject with the
question: can existing photoproduction data be used to test the
octet-mechanism?

\section{Color-Singlet Approaches to Photoproduction of $J/\psi$}

Before discussing the NRQCD factorization formalism, we first discuss
other theoretical frameworks for the calculation of charmonium
production rates. One of the earliest computations of $J/\psi$
production was carried out by Berger and Jones in 1981~\cite{bj}. 
There, the authors give an expression for the rate of production of
$J/\psi$ in $\gamma$-nucleon collisions, as calculated in the
so-called ``color-singlet" model.  In this model, one adopts a
nonrelativistic boundstate picture to describe the
$J/\psi$~\cite{review}; calculations of $J/\psi$ production are based
upon the amplitude for the generation of a $c\bar{c}$ pair in a
color-singlet ${}^3S_1$ configuration with small relative momentum
({\it i.e.} $|{\bf q}| \ll m_{c}$).  The $c\bar{c}$ subprocess
considered by Berger and Jones is shown in Fig.~1.  In it, a gluon
from the nucleon ``fuses'' with the photon to form a hadronizing
$c\bar{c}$ pair  that recoils against a gluon jet. These diagrams
represent the leading color-singlet contribution to
\begin{equation}
\label{mainequation}
\gamma \; + \; N \;\; \rightarrow \;\; J/\psi + X \; .
\end{equation}
The color-singlet model formulation of $J/\psi$ photoproduction has
been studied extensively~\cite{thephotopro}. These analyses hinge on
the experimental kinematic parameter
\begin{equation}
\label{zdef}
z \equiv \frac{P_{J/\psi}\cdot P_{N}}{P_{\gamma}\cdot P_{N}},
\end{equation}
where $P_{J/\psi}$ is the $J/\psi$ four-momentum, $P_N$ is the
four-momentum of the initial state nucleon, and $P_{\gamma}$ is the
initial state photon four-momentum. (Note that in the lab frame of
fixed target experiments, one has $z = E_{\psi}/E_{\gamma}$.)  It is
found that once next-to-leading order QCD corrections to the
Berger-Jones results are taken into account, the color-singlet model
can adequately explain the experimental data for the kinematic regime
$p_T \geq 1$GeV and $z \leq 0.8$, where $p_{T}$ is the transverse
momentum of the $J/\psi$ in the center-of-mass frame.

\begin{figure}
\begin{center}
\epsfxsize=0.5\hsize
\mbox{\epsffile{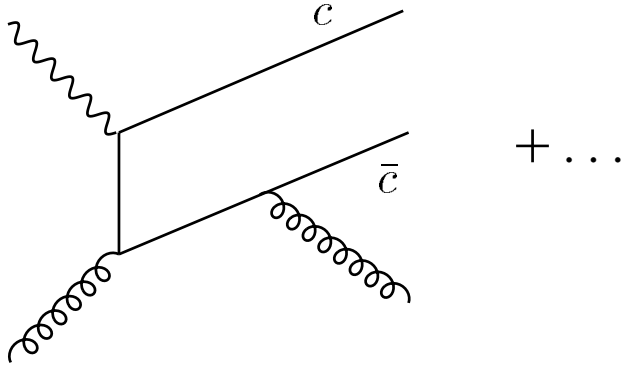}}
\end{center}
\caption{Leading order diagrams for the photoproduction of a
{}$c\bar{c}$ in a color-singlet, ${}^3S_1$ state.}
\end{figure}

However, the color-singlet model alone cannot explain the total
$J/\psi$ photoproduction cross section. It severely underestimates
the rate of production in the region $z \ge 0.9$~\cite{kc}~\cite{h1},
which (importantly) is the region containing nearly all of the
events. In production of $J/\psi$ in this high-$z$ range, the ``$X$''
in Eq.~\ref{mainequation} represents either an elastically scattered
nucleon, or, a nucleon (or resonance thereof) and light hadrons that
have very low energy in the rest frame of the final state nucleon.
Such events, sometimes called ``diffractive scattering,'' have been
modeled in Refs.~\cite{brodsky} and \cite{ryskin}. The essential
features of the diffractive mechanism are the creation of a heavy
quark pair from the incoming photon and the exchange of some
color-singlet gluon combination between the $c\overline{c}$ pair and
the proton.  (Diffractive processes can in general entail the
creation of a heavy quark pair from an incoming gluon also.) In the
analysis of Refs.~\cite{brodsky} and \cite{ryskin}, the amplitude for
the completely exclusive process $\gamma  p  \rightarrow  J\!/\!\psi
\, p$ is factored into three pieces: the process $\gamma \rightarrow
c\bar{c}$, the scattering of the $c\bar{c}$ system on the proton via
(colorless) two (or multiple) gluon exchange, and the formation of
the $J/\psi$ from the outgoing $c\bar{c}$ pair. The exchanged gluons
are taken to have very low transverse momentum.  This mechanism is
considered in Ref.~\cite{ryskin} to serve as a probe of the gluon
density in the proton, the cross-section being proportional to $[x
f_{g/N}(x)]^2$:
\begin{equation}
\label{diffractive}
d \sigma (\gamma p \rightarrow J/\psi p)/dt \propto
\left[ x f_{g/N}(x)\right]^2
\end{equation}
where $x$ is a dimensionless ratio of kinematic variables. Indeed
there is remarkably good agreement between the formula of
Refs.~\cite{brodsky} and \cite{ryskin} and high-$z$ fixed target and
HERA data (see Fig.~2 of Ref.~\cite{ryskin}).

The photoproduction cross section can be described at low $z$ by the
color-singlet model, and at high $z$ by the approach of
Refs.~\cite{brodsky} and \cite{ryskin}. It is interesting however to
consider the possibility that the total $J/\psi$ photoproduction
cross section can be parametrized within a single coherent
theoretical framework. Such a framework is provided by the
nonrelativistic QCD (NRQCD) factorization formalism of Bodwin,
Braaten, and Lepage~\cite{bigbbl}. This formalism is intended only
for the parameterization of inclusive production of quarkonium. 
Here, ``inclusive'' means that a perturbative partonic process
produces the heavy quark pair and (possibly) some other specified
final state partons; the heavy quark pair hadronizes into a specific
quarkonium boundstate but no mention is made in the formalism of the
fate of the other partons regarding the baryons into which they
hadronize; all possibilities are {\it included}; moreover, a large
number of light quanta (generally pions) are expected to accompany
this {\it inclusive} production.

At the outset, the choice between the diffractive scattering
formalism and the NRQCD factorization formalism hinges on the
questions that one asks, that is whether one desires to model
exclusive or inclusive events.  We will see below however that the
line between these two cases is blurred by the fact that in the set
of all data, a large portion of the $J/\psi$ photoproduction events
have a certain exclusive character --- that is, the final state
contains very few of the extra quanta (such as pions) that one would
expect in inclusive production.

\section{NRQCD Factorization Formalism}

According to the NRQCD factorization formalism, the inclusive cross
section for quarkonium production is expressed as a sum of products
known as a ``factorization formula." For the photoproduction of
$J/\psi$, the factorization formula is
\begin{equation}
\label{factorizedform}
\sigma(\gamma + N \to J/\psi + X) = \sum_n {F( n) \over m^{d_{ n}-4}_{c}}
\langle 0|  {\cal O}^{J/\psi}( n)  0|\rangle \; .
\label{ffcs}
\end{equation}
In the above expression, the index $n$ labels the initial color and
angular-momentum quantum numbers of the $c\bar{c}$ pair produced by
short-distance physics. The short-distance coefficients $F(n)$
contain only effects of distance scales of order $1/m_c$ (where $m_c$
is the charm quark mass) or smaller; they can be calculated, using
Feynman diagrams, as a perturbative expansion in $\alpha_s(m_{c})$. 
Effects of longer distances, including effects related to the
hadronization of the $c\bar{c}$ pair into the $J/\psi$ boundstate,
are parametrized by the NRQCD matrix elements $\langle 0| {\cal
O}^{J/\psi}(n) |0 \rangle$; their relative importance can be
determined using the NRQCD $v$-scaling rules~\cite{vscaling}, where
$v$ is the typical relative velocity of the $c$ and $\bar{c}$ in the
$J/\psi$ boundstate. $d_n$ is an integer associated with the energy
dimension of the operator ${\cal O}^{J/\psi}(n)$.

The NRQCD formalism has several features that make it highly
compelling.  Firstly, it is rigorous in that it is based upon an
effective Lagrangian for heavy quarks derived directly from full QCD.
Secondly, it goes beyond the color-singlet model in that it takes
into consideration processes in which a heavy quark pair is produced
initially in a {\it color-octet} state and attains color-neutrality
and hadronizes via the emission and absorption of soft gluons; the
rates due to such color-octet channels are rigorously and
systematically parametrized in terms of NRQCD matrix elements.
Finally, the NRQCD factorization formalism provides a solution to the
problem of the infrared divergences that arise in naive
color-singlet-model calculations of P-wave quarkonia decay and
production~\cite{bbllett}; in the NRQCD framework, these divergences
are absorbed into appropriate (color-octet) matrix elements.

The factorization hypothesis expressed in Eq.~\ref{ffcs} is
considered to be most valid when applied to differential cross
sections with $p_T$ much greater then the hadronic scale $1 \;
\mbox{GeV}$ \cite{css}. However, it is not yet clearly understood in
the NRQCD theoretical community whether {\it total} cross sections
can be expressed in the factorized form. In this paper we will assume
that they can be. If our assumption is valid then one expects
relative corrections from higher-twist operators to Eq.~\ref{ffcs} to
be\cite{gbpc}
\begin{equation}
\label{fundamentallimit}
O\left(  (1 \; \mbox{GeV})^2/ m_c^2 \right) ,
\end{equation}
with $m_c$ in GeV.

\section{NRQCD Prediction for Photoproduction}

Let us now turn to a discussion of the NRQCD prediction for the total
cross section for photoproduction of $J/\psi$. To write down the
leading pieces of the NRQCD prediction, we must first determine which
are the most important terms in the factorization formula. The
numerical sizes of the NRQCD matrix elements can be estimated by
determining how they scale with $v$, which, as stated earlier, is the
typical velocity of the heavy quarks in the quarkonium boundstate. 
($v^2_c \simeq 0.25$.) Combining $v$-scaling estimates of $\langle
0|{\cal O}^{J/\psi}( n) |0\rangle$ with the $\alpha_{s}$-scaling of
the $F( n)$, it is possible to determine the relative importance of
the terms in the factorization formula, Eq.~\ref{ffcs}, in regard to
the double expansion in $v$ and $\alpha_{s}$. One finds that the
leading contributions to the total photoproduction cross section come
from the production of a $c\bar{c}$ pair in a color-singlet
${}^{3}S_{1}$ state (the Feynman diagrams are shown in Fig.~1), and
from the production of a $c\bar{c}$ pair in a color-octet 
${}^{1}S_{0}$, ${}^{3}P_{0}$, or ${}^{3}P_{2}$ state (the Feynman
diagrams are shown in Fig.~2). The leading color-singlet contribution
to the rate is proportional to $\alpha_s^3 v^3$, while the leading
color-octet contributions to the rate are proportional to $\alpha_s^2
v^{7}$. Since $\alpha_s(m_c) \sim v^2$, we expect the two effects to
contribute roughly with the same size to the total cross-section.
\begin{figure}
\begin{center}
\epsfxsize=0.5\hsize
\mbox{\epsffile{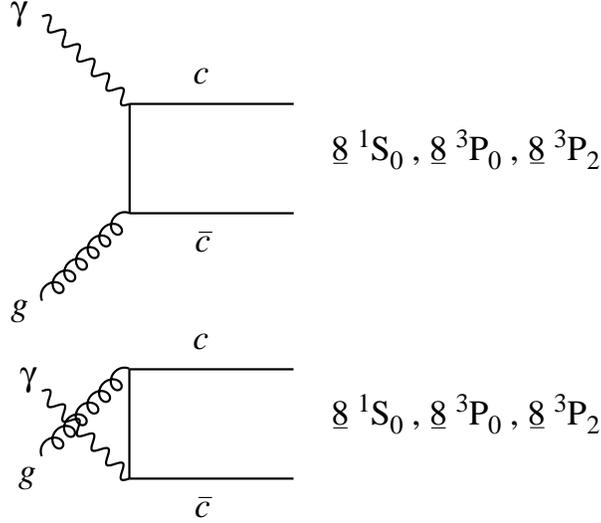}}
\end{center}
\caption{Leading order diagrams for the photoproduction of a
{}$c\bar{c}$ in a color-octet state with angular momentum
configuration ${}^1S_0$, ${}^3P_0$, and ${}^3P_2$.
These diagrams are the underlying subprocess for the
forward octet contributions to $J/\psi$ photoproduction.}
\label{diagrams}
\end{figure}

Note that in the case of $J/\psi$ production where the $c\bar{c}$
pair is produced initially in a color-singlet ${}^{3}S_{1}$ state (as
in Fig.~1), there must necessarily be a final state hard gluon, for
color conservation. If this hard gluon is of relatively low (high)
energy, then the $J/\psi$ is produced in the high-$z$ (low-$z$)
region. For the color-singlet contribution, we thus see that there
exists some continuous $z$-distribution. On the other hand, as to the
case (in Fig.~2) in which the $c\bar{c}$ pair is initially produced
in a color-octet ${}^{1}S_{0}$, ${}^{3}P_{0}$, or ${}^{3}P_{2}$
state,  there is no hard final state gluon at leading order; the
color-octet mechanism contributes therefore only to the high-$z$
region at leading order.  The naive picture is that the
$z$-distribution from this sort of contribution is an idealized Dirac
function at $z = 1$.  We will refer to the color-octet process in
Fig.~2 as ``forward-octet.''

We now present the color-singlet $J/\psi$ photoproduction rate, the
underlying process of which is shown in Fig.~1.  The NRQCD result for
the color-singlet contribution is simply proportional to the
color-singlet model expression~\cite{bj}~\cite{thephotopro}
\begin{equation}
\label{cscontribution}
\frac{d\sigma_{{\rm CS}} }{dt} \;  = \;
\frac{64 \pi e_c^2 \alpha_{em} \alpha_s(t)^2  m_c | R_s(0) |^2}{ 3 s^2} \;
\frac{s^2 (s - 4m^2_c)^2 + t^2 (t - 4m^2_c)^2 + u^2 (u - 4m^2_c)^2}
{ (s - 4m^2_c)^2 (t - 4m^2_c)^2 (u - 4m^2_c)^2 }
\end{equation}
where $\alpha_s(t)$ is the strong coupling constant evaluated at the
typical scale of the interaction, $t$, where $\alpha_{em}$ is the
electromagnetic coupling constant, where $e_c^2$ is the fractional
charm quark charge, and where $s$, $t$, and $u$ are the usual Lorentz
invariant Mandelstam variables. Here $R_s(0)$ is the radial
color-singlet wavefunction evaluated at ${\bf x} = 0$. It is related
to the color-singlet ${}^{3}S_{1}$ production matrix element through
\begin{equation}
\langle 0|{\cal O}_1^{J/\psi}({}^{3}S_{1})|0\rangle =
{9 \over 2 \pi} |R_{s}(0)|^{2} \big( 1 + O(v^{2}) \big) \; .
\label{fftocsm}
\end{equation}

We next present the leading color-octet (``forward octet'')
photoproduction rate, the underlying process of which is shown in
Fig.~2. The factorization formula for the forward-octet contribution
can be written as follows:
\begin{eqnarray}
\sigma(\gamma g \to J/\psi +X)_{{\rm FO}} &
= & {F(8,{}^1S_0) \over m^{2}_{c}}
\langle 0| {\cal O}^{J/\psi}_8 ({}^1S_0) | 0 \rangle
\nonumber\\
& & + {F(8,{}^3P_0)  \over m^{4}_{c}}
\langle 0| {\cal O}^{J/\psi}_8 ({}^3P_0) | 0 \rangle
+ {F(8,{}^3P_2)  \over m^{4}_{c}}
\langle 0| {\cal O}^{J/\psi}_8 ({}^3P_2) | 0 \rangle
\; .
\label{lofacform}
\end{eqnarray}
One can lift (with the help of a suitable color-factor replacement),
the short-distance coefficients $F(8,{}^{1}S_{0})$ and
$F(8,{}^{3}P_{J})$, from the results for the process $gg \to J/\psi$
given in Ref.~\cite{us}.  One obtains
\begin{eqnarray}
F(8,{}^1S_0) & = &
{\pi^3 e^{2}_{c} \alpha_s(2m_{c}) \alpha_{em} \over m_c}
\; \delta(4m^2_c - s)
\nonumber \\
F(8,{}^3P_0) & = &
{3 \pi^3 e^{2}_{c} \alpha_s(2m_{c}) \alpha_{em} \over m_c} \;
\delta(4m^2_c - s)
\nonumber \\
F_8(8,{}^3P_2) & = &
{4 \over 5} {\pi^3 e^{2}_{c} \alpha_s(2m_{c}) \alpha_{em} \over m_c} \;
\delta(4m^2_c - s)  \; .
\label{sdcoef}
\end{eqnarray}
Inserting these into the factorization formula in
Eq.~\ref{lofacform}, we obtain the subprocess cross-section
\begin{equation}
\sigma(\gamma g \to J/\psi +X)_{{\rm FO}} = {\pi^3 e^{2}_{c}
\alpha_s(2m_c) \alpha_{em} \over
m_c^3 } \; \delta(4m^{2}_{c} - s ) \; \Theta \; ,
\label{subprocs}
\end{equation}
where $\Theta$ is given by
\begin{equation}
\Theta \equiv \langle 0|{\cal O}^{J/\psi}_{8}({}^1S_0)|0 \rangle
+ {7\over m_c^2}\langle 0| {\cal O}^{J/\psi}_{8}({}^3P_0)|0\rangle \; .
\end{equation}
The above expression is derived with the help of the
relation~\cite{bigbbl}
\begin{equation}
\langle 0| {\cal O}^{J/\psi}_8 ({}^3P_J) | 0 \rangle =
(2J+1)\langle 0| {\cal O}^{J/\psi}_8 ({}^3P_0) | 0 \rangle
\bigg( 1 + O(v^{2}) \bigg) \; .
\label{jrelation}
\end{equation}

We next convolute the subprocess cross sections given in
Eq.~\ref{cscontribution} (color-singlet) and Eq.~\ref{subprocs}
(forward-octet) with the gluon distribution function to obtain the
leading NRQCD factorization formalism prediction for the {\it total}
photoproduction cross section:
\begin{eqnarray}
\label{parton}
%
%
&& \sigma(\gamma N \to J/\psi + X)  \nonumber \\
&& \qquad = \int dx \; f_{g/N}( x ) \;
\Big( \sigma_{{\rm CS}}(\gamma g \to J/\psi + X) +
\sigma_{{\rm FO}}(\gamma g \to J/\psi + X) \Big) \nonumber\\
&& \qquad  =
{\pi^2 \alpha e^2_c \over m^3_c} \; \int dx \; f_{g/N}( x ) \;
\Bigg(\pi \alpha_s(2m_c)  \; \Theta \; \delta
\left( 4m_c^2- \hat{s} \right) \; \; +
\nonumber \\
&& \qquad \int d\hat{t} \frac{128 \alpha_s^2(\hat{t}) m^4_c
\langle 0| {\cal O}_1^{\psi}({}^{3}S_{1})|0\rangle}{ 27 \hat{s}^2} \;
\frac{\hat{s}^2 (\hat{s} - 4m^2_c)^2 + \hat{t}^2 (\hat{t} - 4m^2_c)^2
+ \hat{u}^2 (\hat{u} - 4m^2_c)^2}
{ (\hat{s} - 4m^2_c)^2 (\hat{t} - 4m^2_c)^2 (\hat{u} - 4m^2_c)^2 }
\Bigg)  \; ,
\end{eqnarray}
where $\hat{s}$, $\hat{t}$, and $\hat{u}$ are the usual Mandelstam
variables for the subprocess cross section, and $f_{g/N}( x )$ is the
gluon structure function.

\section{Comparison to experiment: Caveats}

The leading NRQCD factorization formula for the total photoproduction
cross section (Eq.~\ref{parton}) depends on the two phenomenological
quantities: $\langle 0|{\cal O}^{J/\psi}_{1}({}^{3}S_{1})|0\rangle$,
and $\Theta \equiv \langle 0|{\cal O}^{J/\psi}_{8}({}^1S_0)|0 \rangle
+ {7\over m_c^2}\langle 0| {\cal O}^{J/\psi}_{8}({}^3P_0)|0\rangle$.
The first parameter (the color-singlet matrix element), is well
determined from measurements of the decay rate of $J/\psi$ to two
leptons \cite{bsk}. As to the second parameter (the forward-octet
combination $\Theta$), the matrix elements contained therein are
poorly constrained thus far, and so a testable prediction of the
photoproduction cross section is not yet possible.

Given this state of affairs, we propose to use the factorization
formula in Eq.~\ref{parton} to obtain an estimate of the value of
$\Theta$ via a comparison with fixed target photoproduction data. We
find that a determination of $\Theta$ is complicated by the manner in
which $J/\psi$ photoproduction data is presented in the experimental
literature. Typically, measurements of the {\it total inclusive}
cross section are not made; thus, our Eq.~\ref{parton} --- a total
inclusive cross section --- cannot be associated in an unqualified
straightforward manner with some published empirical result. We must
contend with the fact that data is taken in either the {\it
inelastic} or {\it elastic} regime. The {\it inelastic} regime is
generally considered to be the region where $z$ is below $0.8$ or
$0.9$.  On the other hand, the conventional definition of the {\it
elastic} regime is not at all well established; it is in general
considered to be the region near $z=1$, with the data consisting of
those events in which is detected a muon pair with invariant mass
$M_\psi$ along with hadrons having up to about 5~GeV of additional
energy. To test the forward-octet feature of NRQCD (which is our
goal), it would seem at the outset that a good strategy would be to
select some range of $z$ (from a lower bound around 0.8 up to unity)
for which to calculate the total (color-singlet plus color-octet)
theoretical prediction, and to compare that result to the elastic
data. However the contribution of the color-singlet term is quite
negligible for high $z$, and is actually found to be nearly an order
of magnitude below the data~\cite{kc}~\cite{h1}. The unimportance of
the color-singlet contribution $d\sigma_{{\rm CS}}/dz$ at high $z$
and the vagueness surrounding the correct lower bound of $z$ to use
when integrating over $z$ in order to compare to elastic data suggest
that we might as well neglect the color-singlet contribution
altogether. Why not simply go ahead and compare the elastic data to
the forward-octet contribution? Before we proceed, some reservations
must first be discussed.

A major reservation concerns the issue of ``$z$-smearing.'' Although
we ultimately do not take this stance, it is arguable that one
actually does {\it not} expect the forward-octet piece to contribute
significantly to photoproduction events for which $z$ is nearly
unity. Although the subprocesses described in Fig.~2 appear, at a
first glance, to create $J/\psi$ particles with total energy close to
that of the incoming photon ({\it i.e.} with $z =
(E_{J/\psi})/(E_\gamma) = 1$), this is not exactly the case since the
$c\bar{c}$ system must emit or absorb gluons in order to make the
transition to a color-singlet state and hadronize into a $J/\psi$
boundstate. Such gluons typically have energy and momenta of order
$m_cv^2$ in the rest-frame of the $c\bar{c}$ system \cite{bigbbl}. 
The presence of these gluons implies that the ratio of the energy of
the final charmonium particle to that of the initially produced heavy
quark pair is not exactly unity but, rather, is expected to be $<z> =
1 - O(v^2)$. The $z$-distribution is smeared away from the
idealization of the Dirac delta $\delta(1 - z)$.  Similar reasoning
has been given in \cite{brm}. In fact, even in quarkonium production
events involving color-singlet channels, one still might expect such
a disparity between the energy of the initially produced heavy quark
color-singlet pair and the final boundstate; the heavy quarks emit
pions and other quanta so as to adjust their energy-momentum to that
of heavy quarks in the boundstate.

The above arguments concerning $z$-smearing seem reasonable, but
experimental information challenges them severely. In fact, quite
contrary to what one might expect, it turns out that the great bulk
of photoproduction events actually lie in bins above $z = 0.95$.  For
example, of the approximately 850 $J/\psi$ photoproduction events
collected by the H1 experiment, 700 lie in the bin above
$z=0.95$~\cite{h1}. This includes both the events in which the proton
is reconstructed, and those in which the proton is dissociated. So,
assuming that these events are due to the NRQCD forward octet
mechanism, we must conclude that the smearing of the $z$-distribution
is not a big effect. \footnote{It must be mentioned in passing that
in the exclusive diffractive scattering, as well, one would expect
$z-$-smearing to occur.  The $c$ and $\overline{c}$, which emerge
from the diffractive scattering in a color-singlet state, would have
to readjust their momenta to that of quarks in the quarkonium
boundstate, and this necessarily entails the emission of soft quanta
of momentum of the order of $m v^2$.}

There is yet another reason for which, at the outset, one might have
reservations concerning the fitting of Eq.~\ref{parton} to high-$z$
fixed target elastic data: the question of the inherently inclusive
nature of the NRQCD formalism versus the possible lack of complete
inclusiveness in the data. However, as it happens, the fixed target
data {\it do} appear to be sufficiently inclusive for our purposes.
This is because the H1 data \cite{h1}, as was mentioned just above,
show that most photoproduction events are associated with $z >
0.95$.  If we can count on all high energy ($\sqrt{s} \gg M_{\psi}$)
$J/\psi$ photoproduction to have this feature, then the elastic fixed
target data that we are using should surely include the great
majority of events, and can be considered inclusive.

The above arguments support the decision to simply associate the
high-$z$ fixed target data with the NRQCD forward-octet contribution.

\section{Comparison to Experiment: Determination of $\Theta$}

We now proceed to use experimental data from the E687, NA14, E401,
NMC and E516 experiments to evaluate $\Theta$.  We will then show
that the value we determine in this way is consistent with HERA
(H1~\cite{h1} and ZEUS~\cite{zeus}) data.

In Fig.~3 we display our fit of the forward-octet piece in
Eq.~\ref{parton} to elastic cross-section measurements from the fixed
target Fermilab experiments E687~\cite{e687}, NA14~\cite{na14},
E401~\cite{e401}, NMC~\cite{nmc}, and E516~\cite{e516}. In the same
figure we also show data from the SLAC experiment~\cite{slac}; we do
not however include this data in our fit since it falls in the low
energy ($\sqrt{s} \sim M_{\psi}$) regime. As indicated in Fig.~2, we
have performed our analysis using CTEQ \cite{cteq}, MRS~\cite{mrs},
and GRV~\cite{grv} structure functions.  We have taken $m_c =
1.5$~GeV and $\alpha_s(2m_c) = 0.26$. With this choice of parameters
we find
\begin{equation}
\label{Thetaw}
\Theta = 0.02 \; \mbox{GeV}^3 \; .
\end{equation}
Note that there are large theoretical errors associated with this
determination of $\Theta$. These will be discussed in the next
section.
\begin{figure}
\begin{center}
\epsfxsize=0.8\hsize
\mbox{\epsffile{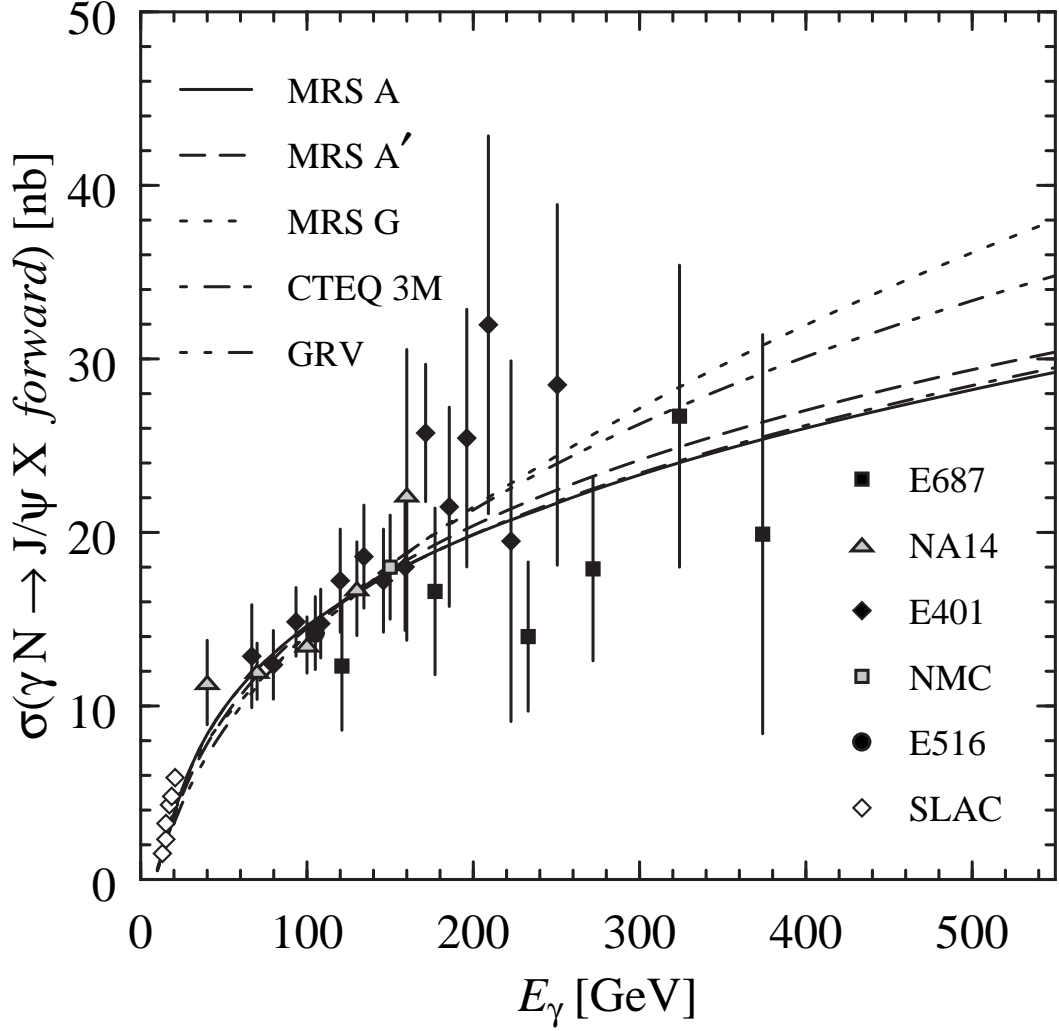}}
\end{center}
\caption{Fit to fixed-target experiments.  Shown are data points for
the elastic photoproduction cross section measured at
fixed-target experiments.  Also shown are the theoretical curves
evaluated numerically using five different gluon distribution
functions.  In the curve $\Theta$ is adjusted to best fit the
data.}
\label{egammafig}
\end{figure}

Having estimated $\Theta$ from fixed target Fermilab data, it is
interesting to see whether we can use this result to make a
successful theoretical prediction for the higher energy elastic data
from HERA. The prediction is shown in Fig.~4, where it is seen that
our extrapolated theoretical curve (generated with $\Theta = 0.02
\mbox{GeV}^3$) is indeed consistent with HERA data. On this plot we
show data from the experiments H1~\cite{h1} and ZEUS~\cite{zeus}.
(Also shown on this plot are measurements from the Fermilab
experiments E401~\cite{e401} and E516~\cite{e516}, which are a class
of data that the HERA collaboration includes on their own plots for
comparison purposes. Note that the E401 and E516 data appearing in
this figure are not the same as those appearing in Fig.~3.)
\begin{figure}
\begin{center}
\epsfxsize=0.8\hsize
\mbox{\epsffile{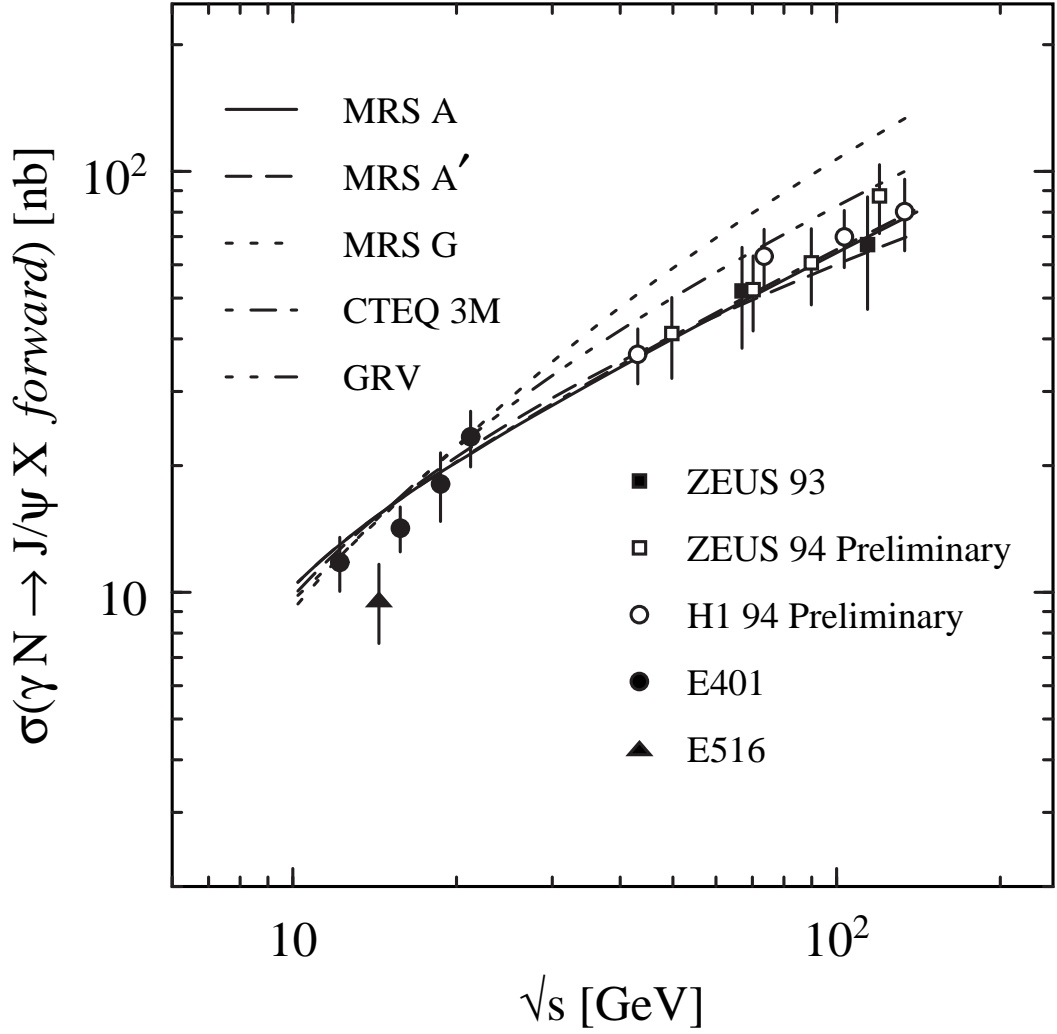}}
\end{center}
\caption{Comparison of NRQCD forward octet theoretical prediction
to HERA data using the value of $\Theta$
determined from fixed target experiments.
Shown are data points for the elastic scattering cross
section measured at various HERA and Fermilab experiments.}
\label{wfig}
\end{figure}

\section{Theoretical Uncertainties}

Here we discuss the sources of theoretical error bearing upon our
determination of $\Theta$.
We have already stated that one expects an error do to the neglect of
``higher-twist'' operators of $O\left(  (1 \; \mbox{GeV})^2/ m_c^2
\right)$. Apart from this limitation, there are other important
sources of theoretical uncertainty. These are associated with
$\alpha_s$ corrections, with relativistic $v^2$ corrections, and with
the value of the parameter $m_{c}$. We now address each of these
issues.

As to $\alpha_s$ and relativistic corrections, to determine their
size, one really must calculate them. When one is not in possession
of such results, the order-of-magnitude of the corrections can
nevertheless be estimated by examining similar processes where
next-to-leading order corrections have been computed. In this matter
we are fortunate, because there exist calculations of the
$\alpha_{s}$ corrections~\cite{thephotopro} and $v^{2}$
corrections~\cite{keung} to the {\it color-singlet} term,
Eq.~\ref{cscontribution}. The extrapolations of these results to a
color-octet production process must be considered no better than 
educated guesses; we seek only order-of-magnitude guidance. According
to Ref.~\cite{thephotopro}, the $\alpha_{s}$ corrections to the
color-singlet piece increase the theoretical result for that part of
the rate by roughly a factor of two; we expect $\alpha_s$
corrections of similar size therefore for the forward-octet terms.
Concerning relativistic corrections on the other hand, the results of
Ref.~\cite{keung} show that $v^2$ corrections to the color-singlet
part of the rate contribute roughly an additional 50\%, and so we
expect the same for the forward-octet part of the rate.

As to the theoretical error due to uncertainty regarding the value of
the parameter $m_c$, we estimate this by varying $m_c$ over the range
1.3 GeV to 1.7 GeV. We find that over this range, the cross section
varies by a factor of $4$.

Clearly, in light of the large size of the various theoretical
uncertainties, the value we determine for $\Theta$ can only be
regarded as an order-of-magnitude estimate.

The uncertainty due to the gluon structure function is negligible in
our analysis. We observe that different standard parameterizations of
the gluon structure function yield results that vary little compared
to the relatively large errors discussed above. Therefore, it is not
possible to make any conclusions about the gluon content of the
nucleon in the present context.

\section{Comparison to Other Work}

We now confront other measurements of the NRQCD $J/\psi$ production
matrix elements with our measurement of $\Theta$.

Beneke and Rothstein~\cite{brm} calculate the leading terms in the
NRQCD factorization formula for $\sigma(\pi N \to J/\psi + X)$. This
hadroproduction reaction involves --- among other processes --- the
forward octet process in Fig.~2, but with two incoming gluons instead
of the photon and gluon as in photoproduction. As a consequence,
$\langle 0|{\cal O}_8^{J/\psi}(^1S_0)|0\rangle$ and $\langle 0|{\cal
O}_8^{J/\psi}(^3P_0)|0\rangle$ appear in hadroproduction rates in the
exact same linear combination ($\Theta$) as they do in
photoproduction. Comparing their factorization formula to data,
Beneke and Rothstein obtain $\Theta = 0.03 \; \mbox{GeV}^3$. This is
consistent with our result, given the large theoretical
uncertainties.

A different linear combination of $\langle 0|{\cal
O}_8^{J/\psi}(^1S_0)|0\rangle$ and $\langle 0|{\cal
O}_8^{J/\psi}(^3P_0)|0\rangle$ is determined from an analysis of
$J/\psi$ production at the Tevatron~\cite{cl}. A fit of theory to CDF
data for $J/\psi$ produced at moderate $p_T$ gives
\begin{equation}
\langle 0|{\cal O}_8^{J/\psi}(^1S_0)|0\rangle
+{3.5 \over m^2_c}\langle 0|{\cal O}_8^{J/\psi}(^3P_0)|0\rangle
= 4.38\pm1.15_{-0.74}^{+1.52}  \; \times 10^{-2} \; \mbox{GeV}^3 \; .
\label{clresult}
\end{equation}
Here we have quoted the fit done by Beneke and Kr\"amer using the
CTEQ4L structure function. Again, within the theoretical uncertainty,
this result is consistent with ours.

It is interesting to consider the possibility that our value for
$\Theta$ (determined from photoproduction) is accurate, and that the
measurement given in Eq.~\ref{clresult} (determined at CDF) is also
accurate. Upon solving Eqs.~\ref{Thetaw} and \ref{clresult}, one
finds that $\langle 0|{\cal O}_8^{J/\psi}(^3P_0)|0\rangle$ is
negative.  Though this may at first appear to be inconsistent and
even meaningless, it is actually not unreasonable, if one considers
that the matrix elements determined here are renormalized matrix
elements~\cite{negme}.

The {\it elastic} photoproduction calculation carried out in the
present paper has also been carried out by Cacciari and
Kr\"amer~\cite{kc}. The expression they obtain for the forward-octet
terms is in agreement with our Eq.~\ref{subprocs}. However, in their
analysis, these authors arrive at conclusions different from ours.
Cacciari and Kr\"amer assume that the production matrix elements must
be positive. They use the result of the analysis of Cho and
Leibovich~\cite{cl} (which is similar to the Beneke and Kramer
result, given here in Eq.~\ref{clresult}), to estimate the order of
magnitude of $\langle 0|{\cal O}_8^{J/\psi}(^1S_0)|0\rangle$ and of
$\langle 0|{\cal O}_8^{J/\psi}(^3P_0)|0\rangle$. They find that,
under these assumptions, the forward-octet term in the theoretical
prediction for the low-$p_T$ high-$z$ cross section is one order of
magnitude greater than experiment. Their conclusion is that the NRQCD
factorization formalism might be in conflict with the data.  This
potential dilemma is alleviated, though, by the fact that, as is
pointed out above, not all the matrix elements need be positive.

Cacciari and Kr\"amer also calculate the color-octet contribution to
the {\it inelastic} (moderate $z$) photoproduction cross section
$d\sigma/dz$. (We have not computed the color-octet contribution to
this regime of the differential cross-section.) Cacciari and Kr\"amer
conclude once more that the NRQCD factorization formalism appears at
variance with data.  In this instance, not only is the theoretical
prediction several times greater than the data, but, this time, the
result of the computation of the cross-section (here, the inelastic
color-octet cross-section) {\it cannot} be affected appreciably by
the fact that $\langle 0|{\cal O}_8^{J/\psi}(^3P_0)|0\rangle$ can be
negative~\cite{kcp}. Nevertheless, it is clear from the long list of
sources of theoretical uncertainty that no strong conclusions can be
drawn from this discrepancy regarding the correctness of the NRQCD
factorization formalism.

\section{Conclusions}

We have calculated the leading order (in coupling constants and $v$)
color-singlet and color-octet contributions to the NRQCD
factorization formula for the photoproduction of $J/\psi$. At this
order the total cross section depends on the two phenomenological
parameters: $\langle 0|{\cal O}^{J/\psi}_{1}({}^{3}S_{1})|0\rangle$,
and $\Theta \equiv \langle 0|{\cal O}^{J/\psi}_{8}({}^1S_0)|0 \rangle
+ {7\over m_c^2}\langle 0| {\cal O}^{J/\psi}_{8}({}^3P_0)|0\rangle$.
Though the color-singlet matrix element is well determined, the
color-octet matrix elements contained in the forward-octet
combination $\Theta$ are poorly constrained. Therefore a testable
prediction of the photoproduction cross section cannot be generated.

Be that as it may, it is nonetheless possible to use fixed target
elastic photoproduction data to obtain a rough estimate of $\Theta$.
There are some reservations to this approach, though. Firstly, there
is the issue of the ``z-smearing'' of the forward-octet contribution.
One might be concerned that the forward-octet contribution is
centered around $< z > = 1 - O(v^2)$, not $z=1$. Secondly, there is
the issue of inclusiveness. NRQCD factorization formulas apply to
inclusive production, and it would seem that the region $z > 0.95$ is
not inclusive enough. However both of these arguments are challenged
by experimental data collected by the H1 collaboration, which shows
that of 850 $J/\psi$ photoproduction events $700$ are contained in
the region $z > 0.95$.

Proceeding with a fit of the forward-octet contribution to fixed
target $J/\psi$ elastic photoproduction data we determine
\begin{equation}
\Theta = 0.02 \; \mbox{GeV}^3.
\nonumber
\end{equation}
Given the large theoretical uncertainties associated with this
calculation, it is best to regard this determination as an order of
magnitude estimate. We also show that, using the value of $\Theta$
quoted above, we can explain the higher energy HERA data. Furthermore
the value we determine for $\Theta$ is consistent with other
determinations of the color-octet matrix elements from hadronic
$J/\psi$ production at fixed target and collider experiments.
Moreover, our value of $\Theta$ is consistent with the $v$-scaling
rules of NRQCD.

It must be appreciated that our NRQCD forward-octet result for the
photoproduction rate in linear is the gluon structure function of the
proton (see Eq.~\ref{parton}) while the diffractive rate calculated
in Ref.~\cite{brodsky} and \cite{ryskin} is quadratic in the gluon
structure function (see Eq.~\ref{diffractive}). Thus, the two methods
are essentially irreconcilable. However, one observes that both
approaches are capable of convincingly describing data.  The extent
to which one might ``prefer'' NRQCD over diffractive scattering as
the correct explanation hinges on whether one accepts the hypothesis
that, when calculating the production rate for heavy quarks, to avoid
double-counting one ought not include certain additional mechanisms
that enhance cross-sections such as a diffractive scattering
\cite{css}.

We have failed to resolve this issue.  Perhaps a valuable diagnostic
test to probe the photoproduction mechanism would be to compute
various polarization rates in both formalisms, for confrontation with
experiment. This includes polarization of the incoming photon, as
well as of the final $J/\psi$. Also, a valuable diagnostic would be
to compare data to various differential cross-sections calculated in
both formalisms.

\medskip
\bigskip
\bigskip
\centerline{{\bf ACKNOWLEDGEMENTS}}

\bigskip

The authors wish to thank Geoff Bodwin, Eric Braaten, and Peter
Lepage for helpful conversations.  The work of I.M. was supported by
the Robert A. Welch Foundation, NSF Grant PHY90-09850, and by NSERC
of Canada.  The work of J.A and S.F. was supported in part by the
University of Wisconsin Research Committee with funds granted by the
Wisconsin Alumni Research Foundation, and by the U.S.\ Department of
Energy under grant DE-FG02-95ER40896.

\newpage

\end{document}